\documentstyle[12pt]{article} 
\begin{document} 
\title{The electric dipole moment of the neutron 
in the constrained minimally supersymmetric standard model}

\author{S.~A.~Abel$^{1}$\thanks{Address after Oct 95: Service de
    Physique Theorique, ULB CP 228, B-1050 Bruxelles}, 
\underline{W.~N.~Cottingham}$^{2}$\thanks{Fax: +44 117 9255624; 
E-mail address: WNC@SIVA.BRIS.AC.UK} \,and 
I.~B.~Whittingham$^{2}$\thanks{Permanent address: 
Physics Department, James Cook University, Townsville, Australia, 
4811}
\\ \vspace{0.3cm} 
\\$^{1}$ Rutherford-Appleton Laboratory, Chilton, Didcot, Oxon OX11 0QX, UK
\\ \vspace{0.3cm}
\\$^{2}$ H H Wills Physics Laboratory, Royal Fort, Tyndall 
Ave, Bristol BS8 1TL UK} 

\maketitle

\begin{abstract} 
We analyse the electric dipole moment of the neutron in the MSSM, induced by
the renormalisation of the soft-susy breaking terms. 
We run the RGEs using two-loop expressions for gauge and Yukawa couplings 
and retaining family dependence. The $\mu$ and $B$ parameters were 
determined by minimising the full one-loop Higgs potential, and we find 
that the neutron EDM lies in the range $10^{-33}<|d_n| < 10^{-29} 
{\; e\;\mbox{cm}}  $.
\end{abstract} 
\vspace*{1 cm}

\newpage

Supersymmetric unified theories are the most promising candidates 
for physics beyond the standard model in that they resolve the 
crucial gauge hierarchy problem of widely separated electroweak and 
grand unified scales, are the consequence of string theories and are 
favoured over non-supersymmetric unified theories by recent high 
precision measurements at LEP. In addition to the usual signatures 
of grand unification such as proton decay, neutrino masses, fermion 
mass relations and weak mixing angle prediction, supersymmetric 
unification is characterised by the resultant mass spectrum of the 
supersymmetric particles (squarks, sleptons, charginos, neutralinos) 
and the flavour-changing and CP-violating processes which arise as 
the renormalisation group equations (RGE) scale the physics from the 
unification scale $M_U \sim 10^{16}$ \mbox{ GeV} down to the electroweak 
scale. Of particular interest are the flavour changing neutral 
current transitions involving the quark-squark-gluino vertex, with
their implications for rare B-decays and mixings \cite{BBM}, and 
the non-removable CP-violating phases, with their implications for 
quark electric dipole moments (EDM) \cite{EDM823,BV94,INUI95} and 
for non-standard-model patterns of CP-violation in neutral B decays 
\cite{BG91}, which result from the RGE scaling of the soft 
supersymmetry (SUSY) breaking scalar interactions in these models.

CP-violation in the standard model (SM) arises from the single phase 
$\delta_{CKM}$ in the Cabbibo-Kobayashi-Maskawa (CKM) mixing matrix
relating the quark weak interaction and mass eigenstates and, in 
principle, this source of CP violation can accommodate the known CP 
properties of the kaon system. Non-zero quark (and lepton) electric 
dipole moments are very sensitive probes of CP violation beyond the 
standard model \cite{DH95} because, unlike the other observables 
of CP violation which are small because of the intergenerational 
mixing angles of the CKM matrix, electric dipole moments are 
{\em particularly} suppressed by the chiral nature of the weak interaction 
and vanish at both one- and two-loop order in the Standard Model, 
resulting in quark EDMs of
\begin{equation}
\label{edm-1}
d^{SM}_{u,d} \sim O(10^{-34})\quad {e\;\mbox{cm}}  .
\end{equation}
At present the experimental bounds on quark EDMs are obtained 
indirectly from measurements of the neutron EDM. In the 
non-relativistic quark model the neutron EDM is
\begin{equation}
\label{edm-6}
d_{n}(quarks) = \frac{4}{3}d_d - \frac{1}{3}d_u
\end{equation}
and is of the same order as the $u$ and $d$-quark EDMs. However the neutron EDM 
is expected to be dominated by long distance effects such that
\begin{equation}
\label{edm-7}
d^{SM}_{n}(LD) = O(10^{-32}) \quad {e\;\mbox{cm}}  .
\end{equation}  

By contrast, non-zero supersymmetric phases, collectively denoted 
$\delta^{SUSY}_{soft}$, arising from the complexity of the soft SUSY 
breaking terms can generate quark EDMs at one-loop order, 
irrespective of generation mixing \cite{EDM823} from diagrams 
involving gluino, chargino (or neutralino) exchange and mixing of 
right- and left-handed virtual squarks. However, for squark, gluino 
and chargino masses of order 100 \mbox{ GeV}, these induced one-loop quark 
EDMs yield a neutron EDM which exceeds the experimental upper bound 
of 
\begin{equation}
\label{edm-2}
d^{EXP}_{n} < 12 \times 10^{-26} \quad {e\;\mbox{cm}} 
\end{equation}
unless these soft phases are constrained to satisfy
\begin{equation}
\label{edm-3}
\delta ^{SUSY}_{soft} < 0.01  .
\end{equation}
The alternative scenario of not imposing any condition of smallness
on these phases but instead making the supersymmetric scalar masses
heavy enough ( of the order of 1 TeV) to suppress the EDMs has been
considered by Kizukuri and Oshimo\cite{KO92}. This scenario also 
has consequences for the relic density of the lightest supersymmetric
particle\cite{FOS95}.

For supersymmetric unified models such as spontaneously broken $N=1$ 
supergravity with flat K\"{a}hler metrics \cite{SUGRA} the 
resultant explicit soft SUSY breaking terms at the scale $M_{SUSY} 
\sim M_U$ of local SUSY breaking are quite simple and these generic 
phases $\delta^{SUSY}_{soft}$ are reduced to just two phases 
$\delta^{SUSY}_{A,B}$ in addition to the usual $\delta_{CKM}$. The 
most natural way of satisfying the experimental bound (\ref{edm-2}) 
and ensuring that the squark, gluino and chargino masses are not 
much above the electroweak scale is to assume that these phases 
$\delta^{SUSY}_{A,B}$ vanish identically at the unification scale 
because of CP conservation in the SUSY breaking sector. Under these 
conditions the only explicit CP violation at the unification scale 
is in the flavour-dependent Yukawa coupling matrices which are 
required to have the structure necessary to reproduce the CKM mixing 
matrix at the electroweak scale under RGE scaling. However, as the 
RGEs for the soft SUSY breaking trilinear couplings $A_{u,d,e}$ and 
bilinear coupling $B$ depend on the Yukawa couplings, inclusion of 
flavour mixing in the RGEs can lead to large RGE-induced 
CP-violating phases in the off-diagonal components of the couplings 
triggered, in particular, by the complexity of the large t-quark 
Yukawa coupling.

The implications of such large phases for EDMs of quarks have been 
studied recently by Bertolini and Vissani \cite{BV94} and Inui 
{\em et al} \cite{INUI95} within a $N=1$ supergravity inspired 
minimally supersymmetric standard model (MSSM) in which the 
spontaneous breaking of the electroweak $SU(2)\times U(1)$ symmetry is 
driven by radiative corrections. Bertolini and Vissani argue that 
the dominant induced EDM is that of the $d$-quark arising from the 
one-loop diagram involving chargino exchange and find
\begin{equation}
\label{edm-4}
d^{SUSY}_{d}(BV) \sim {\cal{O}}(10^{-30})\quad {e\;\mbox{cm}}  ,
\end{equation}
four orders of magnitude greater than the standard model prediction 
(\ref{edm-1}) but still satisfying the experimental upper bound 
(\ref{edm-2}). Inui {\em et al} also find that the $d$-quark EDM 
from chargino exchange is dominant but obtain the much larger value
\begin{equation}
\label{edm-5}
d^{SUSY}_{d}(INUI) \sim O(10^{-27} - 10^{-29})\quad {e\;\mbox{cm}}
\end{equation}
which they ascribe to the inclusion of gaugino masses in their RGEs.

Recently Dimopoulos and Hall \cite{DH95} have considered quark and 
lepton EDMs in a class of supersymmetric unified theories based on 
the gauge group SO(10) where the unification of all quarks and 
leptons of a particular generation into a single {\bf 16} spinorial 
representation leads to non-removable CKM-like phases in the Yukawa 
couplings which, under RGE scalings induced by a large $t$-quark 
Yukawa coupling, give rise to EDMs close to the experimental limits 
such that some regions of the parameter space of the minimal SO(10) 
theory are excluded.

The sensitivity of the quark EDMs to CP violation, and the fact that  the
EDMs of the Standard Model, the MSSM and GUT theories are nicely separated 
makes this an important window to physics at the unification scale. In addition
the dipole moments in the MSSM are a minimum prediction  of supersymmetry.
Because of this, and also because of the numerical discrepancies between the
calculations of Bertolini and Vissani and Inui {\em et al} for the EDM of the
$d$-quark in the constrained MSSM, and the limited nature of the free
parameters chosen in both sets of calculations, we have  undertaken a more
detailed study of quark EDMs.
                                                            
We have used two loop evaluation of gauge and Yukawa couplings, rather than the
one-loop RGEs as used for the existing  EDM calculations, and minimised the
full one-loop Higgs potential,  including contributions from matter and gauge
sectors. We do this following the very complete analyses of Kane {\em et al}
\cite{kane} and Barger {\em et al} \cite{BBO94}, but retaining the full flavour
dependence in the RGEs  as we run them. We do not consider it necessary to
describe the entire procedure since this is outlined in some detail in
refs.\cite{BBO94,kane}, but we shall briefly consider some details of our
analysis.
                                              
The superpotential of the MSSM is given by 
\begin{equation}                                               
W=h_u Q^\dagger_L H_2 U_R + h_d Q^\dagger_L H_1 D_R
 + h_e L^\dagger H_1 E_R + \mu H_1 \epsilon H_2,
\end{equation}    
where generation indices are implied, and we define the VEVs of the  Higgs
fields ($v_1$ and $v_2$) such that $m_u = h_u v_2$,  $m_d = h_d v_1$ and 
$m_e =h_e v_1$. Supersymmetry may be broken softly  by generic mass-squared
scalar terms, gaugino masses, and by `trilinear' couplings  of the form, 
\begin{equation}                                               
\delta {\cal{L}}=A_u h_u {\tilde q}_L h_2 {\tilde u}_R 
+ A_d h_d {\tilde q}_L h_1 {\tilde d}_R
+ A_e h_e {\tilde l} h_1 {\tilde e}_R + B \mu h_1 \epsilon h_2,
\end{equation}    
where again, generation indices are suppressed. In order to determine the
Yukawa couplings at the weak scale, we first ran the Standard Model down to $1
\mbox{ GeV}$ using the two-loop QED   and three-loop QCD RGEs of Arason {\em
et al} \cite{arason}. We then ran  the full supersymmetry RGEs up to the GUT
scale (i.e. where the $SU(2)$  and $U(1)$ gauge couplings unified). For this we
used two-loop RGEs for the  gauge and Yukawa couplings \cite{BBO93},  and
one-loop RGEs for everything else, in the $\overline{DR}$ scheme.  Here we
`unify' by setting the strong coupling equal to the unified $SU(2)$ and $U(1)$
couplings.  This neglects the effects of thresholds at the GUT scale which
depend on the precise details of the GUT theory, and tends to give values of
$\alpha_s$ ($\approx 0.126$ for a top mass of $174 \mbox{ GeV}$) at the weak
scale  which are a little on the high side \cite{kane2}. At the GUT scale there 
are three parameters which we set by hand, the common scalar mass $m_0$, the
common gaugino mass $m_{1/2}$, and the common trilinear coupling $A$. Some of
this degeneracy (for example of the gaugino masses) is  motivated by the
presumed existence of a GUT theory, and some by  minimal supergravity (together
with the assumption that the effects of renormalisation between the GUT and
Planck scales is small).  From the point of view of determining the effects of
renormalisation  of the pure MSSM on the electric dipole moments, degeneracy of
these parameters  is the natural assumption. We then run the entire theory back
down to the  weak scale. We determined the mass eigenvalues and eigenstates  at
the relevant physical scale, $Q=m(Q)$.  At the same time we retained the full 
supersymmetric spectrum in the {\em running} theory, and only decoupled states
in the running of the gauge and Yukawa couplings. Given the value of $\tan
\beta$  and the sign of $\mu$, it is possible to minimise the effective
potential  using the tadpole equations of ref.\cite{BBO94}  in the running
theory. This is a valid procedure if we  wish to minimise the one-loop
effective potential only, and avoids the need to match Lagrangians at each
particle threshold  \cite{kane,BBO94}.  The minimisation was done at
$Q=m_{top}$ (which was taken to be $174 \mbox{ GeV}$ throughout), leaving the 
canonical, hybrid, four-dimensional, parameter space ($m_0$, $m_{1/2}$,  $A$,
$\tan\beta$) in addition to sign($\mu$). However, since all the contributions
from matter and gauge sectors were included in the minimisation, the vacuum
expectation  values of the two neutral Higgs fields (or equivalently the values
obtained for $B$ and $\mu$) should be insensitive to  the momentum scale at
which they are evaluated \cite{GRZ90}. This was indeed found to be the case.
The whole process was then iterated a number of times. Generally, the procedure
converges very rapidly (within a few iterations), and we accurately recover the
entire spectrum given in ref.\cite{BBO94} for a  wide range of
parameters\footnote{We would like to thank P.~Ohmann for  discussions}.
                                        
The mass matrices (in the super-KM basis) were diagonalised numerically to
yield the required mass eigenstates and diagonalisation matrices as follows, 
\begin{eqnarray}
&\mbox{squarks : }& V_{\tilde{q}}^\dagger M^2_{\tilde{q}} V_{\tilde{q}} 
= m^2_{\tilde{q}}\nonumber\\
&\mbox{neutralinos : }& V_N^\dagger M_{N} V_N = m_{\chi^0}\nonumber\\
&\mbox{charginos : }& U_C^\dagger M_C V_C = m_{\chi^{\pm}},
\end{eqnarray}
and the following constraints applied. 

In addition to experimental constraints (we adopt those used in 
ref.\cite{kane}, $m_{\chi^0} > 18 \mbox{ GeV}$, $m_{\chi^\pm} > 47 \mbox{
GeV}$,  $m_{h^0}>44 \mbox{ GeV}$, $m_{h^{\pm}}>44 \mbox{ GeV}$, $m_{A}>21
\mbox{ GeV}$, $m_{\tilde{g}}>141 \mbox{ GeV}$, $m_{\tilde{\nu}}>43 \mbox{
GeV}$, $m_{\tilde{q}}>45 \mbox{ GeV}$)  we insisted that the minimum was stable 
in the sense that the Higgs and squark mass-squareds were positive. We also
required  that the minimum which we obtained was global, and that there were 
no other minima which may have broken colour or charge. The constraints
\begin{eqnarray}
A_\tau^2 & < & 3 h_\tau^2 (m_{\tilde \tau}^2 + m_L^2 + m_1^2)\nonumber\\
A_b^2    & < & 3 h_b^2    (m_{\tilde b}^2 + m_Q^2 + m_1^2)\nonumber\\
A_t^2    & < & 3 h_t^2    (m_{\tilde t}^2 + m_Q^2 + m_2^2)
\end{eqnarray}
provide a coarse indication of this \cite{color}. We also insisted that the
lightest supersymmetric partner was the neutralino, and finally we  required
that the process converged (i.e. that our choice of parameters was not  too
close to any fixed points).     
                        
The diagonalisation matrices appear in trilinear couplings between the quarks
and the heavy supersymmetric scalar bosons and fermions, in particular squarks
and charginos or gluinos. It is CP violation  (i.e. non-zero phases) in these
matrices, at the interaction vertices of the diagrams shown in figs(1a,1b,2),
which may induce a non-zero EDM.  As discussed in ref.\cite{INUI95}, we find
that such phases are indeed induced  into the $A$-terms by the running of the
RGEs, and hence  into the diagonalisation matrices.

Having established this fact, let us consider how the EDM arises. For
completeness, we wish to include the $u$-quark contribution (which is usually
neglected), and so we shall briefly re-examine the EDM calculation.  In doing
so we also hope to gain a little insight into the CP  violating nature of the
EDM. First focus on a gaugeless Lagrangian with  two fermionic fields, one
scalar field, and a single cubic coupling,            
\begin{eqnarray}
\cal{L}&=&\overline{\psi}_1 \left(i\gamma^\mu\partial_\mu-m_1\right)\psi_{1}
        +\overline{\psi}_2 \left(i\gamma^\mu\partial_\mu-m_2\right)\psi_{2}   
\nonumber\\                                                            
       & & + \left|\partial_\mu \phi\right|^2 - m_\phi^2 \left|\phi\right|^2
\nonumber\\                                                            
       & & + \left( a \overline{\psi}_{2L}{\psi}_{1R} +
b \overline{\psi}_{2R}{\psi}_{1L} \right)\phi^* + \left( a^*
\overline{\psi}_{1R}{\psi}_{2L} +
b^* \overline{\psi}_{1L}{\psi}_{2R} \right)\phi .
\end{eqnarray}                                              
The $\psi_1$ field is the light quark whose EDM we would like to calculate, the
scalar field, $\phi$, represents the squark fields, and $\psi_2$ is the  heavy
fermion field (not another quark). For any particular quark, we may choose that
basis in which the mass parameters are real. A CP transformation on this
Lagrangian shows that, in order to have  CP violation, the phases of $a$ and
$b$ must be different (a common phase may be absorbed into the definition of
$\phi$).  Now consider the self energy graphs in fig(1a,b).  In addition to
giving mass and wave function renormalisation,  these diagrams also induce
non-local terms which may be obtained  by performing a derivative expansion     
\begin{eqnarray}
\Delta\cal{L}&=&D\left(\partial_\mu\overline{\psi}_{1R}\partial^\mu
{\psi}_{1L}-m_1^2\overline{\psi}_{1R}{\psi}_{1L}\right) G(x) \nonumber\\
             &+&D^*\left(\partial_\mu\overline{\psi}_{1L}\partial^\mu
{\psi}_{1R}-m_1^2\overline{\psi}_{1L}{\psi}_{1R}\right) G(x)+\ldots
\end{eqnarray}
where the dots represent terms which are higher order in momentum, 
and where 
\begin{eqnarray}
G(x)&=&\frac{1}{(1-x)^3}\left(1-x^2+2 x \log x\right)\nonumber\\
D   &=&  a^* b/\left(32\pi^2 m_2\right)\nonumber\\
x   &=&  m_{\phi}^2/{m_2^2} .
\end{eqnarray}
With CP violation $D$ is complex. The EDM appears when we now introduce 
electromagnetic interactions whilst keeping this expansion 
gauge invariant by introducing covariant derivatives, 
\begin{eqnarray}
\partial_\mu\psi_1 &\rightarrow& (\partial_\mu+i q_1 A_\mu)\psi_1
\nonumber\\
\partial_\mu\psi_2 &\rightarrow& (\partial_\mu+i q_2 A_\mu)\psi_2
\nonumber\\
\partial_\mu\phi   &\rightarrow& (\partial_\mu+i q_\phi A_\mu)\phi
\end{eqnarray}
with $q_1=q_2+q_\phi$. When $D$ is complex, one can anticipate an electric
dipole moment from $\Delta\cal{L}$, of 
\begin{equation}
d=q_1 G(x) \mbox{Im}(D).
\end{equation} 
In fact, when the heavy fermion is electrically neutral (like the gluino of
fig(1a)), this {\em is} the one-loop contribution to the dipole moment.  In
general there is an additional gauge invariant contribution to the  Lagrangian,
coming from the heavy fermion charge. To determine this we must  resort to the
usual one-loop diagram shown in fig(2). It is found to be of the form 
\begin{equation}
\Delta{\cal{L}}' = -q_2 F_{\mu\nu}H(x)\left(
 \mbox{Re}(D) \overline{\psi}_{1}\sigma^{\mu\nu}{\psi}_{1}-
i\mbox{Im}(D)\overline{\psi}_{1}\sigma^{\mu\nu}\gamma^5{\psi}_{1}\right),
\end{equation}
where,
\begin{eqnarray}
F_{\mu\nu}     &=&\partial_\mu A_\nu -\partial_\nu A_\mu,\nonumber\\
\sigma^{\mu\nu}&=&\frac{i}{2} [\gamma^\mu,\gamma^\nu ],
\end{eqnarray}
and where,
\begin{equation}
H(x)=\frac{2}{(1-x)^2}\left(1-x+x\log x\right).
\end{equation}
The total EDM of a quark coming from chargino/squark loops is then,
\begin{equation}
d=\mbox{Im}(D)\left[ q_1 G(x)-q_2 H(x) \right]. 
\end{equation}
For each of the quarks this gives,
\begin{eqnarray}
d_d&=&\frac{e}{3} \mbox{Im}(D_d)F_d(x) \nonumber\\
d_u&=&-\frac{2 e}{3}\mbox{Im}(D_u)F_u(x),
\end{eqnarray}                        
where we have defined the functions, 
\begin{eqnarray}
F_d&=&\frac{1}{(1-x)^3}
\left[ 5-12x+7x^2+2x(2-3x)\log x\right] \nonumber\\
F_u&=&\frac{1}{(1-x)^3}
\left[ 2-6x+4x^2+x(1-3x)\log x\right].
\end{eqnarray}                        
The above analysis generalises in a straightforward manner. The coupling
constants $a$ and $b$ become matrices $a_{ij}$ and $b_{ij}$, with 
$i,j$ running over the appropriate mass eigenstates. 
For the chargino contributions we find 
\begin{eqnarray}
d_d&=&-\frac{1}{3}\frac{e}{32 \pi^2} 
\sum_i^2 
\frac{(V_C)_{2i}(U_C)_{2i}}{m_{\chi_i^\pm}}
\mbox{Im}
\left(h_u \left[ V_{\tilde{u}} 
F_d\left(\frac{m_{\tilde{u}}^2}{m_{\chi_i^\pm}^2}\right)
 V_{\tilde{u}}^\dagger \right]^T_{RL} K^\dagger h_d\right)_{11}\nonumber\\
d_u&=&\frac{2}{3}\frac{ e}{32 \pi^2}  
\sum_i^2 
\frac{(V_C)_{2i}(U_C)_{2i}}{m_{\chi_i^\pm}}
\mbox{Im}
\left(K^\dagger h_d \left[ V_{\tilde{d}} 
F_u\left(\frac{m_{\tilde{d}}^2}{m_{\chi_i^\pm}^2}\right)
 V_{\tilde{d}}^\dagger \right]^T_{RL} h_u \right)_{11},
\end{eqnarray}                        
where we are using the down-quark diagonal basis, and where
$K$ is the CKM matrix. For the gluino contributions we find, 
\begin{eqnarray}
d_d&=&-\frac{e \alpha_s}{9 \pi m_{\tilde{g}}} 
\mbox{Im}\left(
\left[ V_{\tilde{d}}
G\left(\frac{m_{\tilde{d}}^2}{m_{\tilde{g}}^2}\right)
 V_{\tilde{d}}^\dagger \right]_{LR}\right)_{11}\nonumber\\
d_u&=&\frac{2 e \alpha_s}{9 \pi m_{\tilde{g}}} 
\mbox{Im}\left(
\left[ V_{\tilde{u}}
G\left(\frac{m_{\tilde{u}}^2}{m_{\tilde{g}}^2}\right)
 V_{\tilde{u}}^\dagger \right]_{LR}\right)_{11}.
\end{eqnarray}                        
In order to present our results, we choose points generated
at random in the parameter space given by 
\begin{eqnarray}
0     < &m_0       & <1 \mbox{ TeV}\nonumber\\
0     < &m_{1/2}   & <1 \mbox{ TeV}\nonumber\\
-1    < &A         & <1 \mbox{ TeV}\nonumber\\
0     < &\tan\beta & <20.
\end{eqnarray}
In practice, values higher than these seldom satisfy all the criteria 
detailed above (i.e. they imply fine-tuning). 
The region below the low $\tan\beta$ fixed point is 
excluded. For a top-quark of mass $174 \mbox{ GeV}$, $A=0\mbox{ GeV}$ 
and $m_{1/2}=m_0=150 \mbox{ GeV}$, this was found to be at $\tan\beta = 2.2$. 

The modulus of the neutron EDM is plotted against $\tan\beta$ in
fig(3). There is a slight tendency for it to be positive, and the
largest values occur for negative $\mu$ and positive $A$. Clearly  the value of
$\tan\beta$ dominates the EDM of the neutron, and we see the approximately
linear behaviour for large $\tan\beta$ coming from the  increased down-quark
Yukawa couplings. Phases feed into the off-diagonal  elements of $A_d$,
especially into $A_{d13}$. The EDM becomes smaller  as we approach the low
$\tan\beta$ fixed point, since the top Yukawa  coupling dominates the running.
Here the gluino contribution to the up-quark can be the dominant contribution.
Elsewhere however, the down-quark,  chargino diagram is nearly always dominant.
No obvious pattern emerges with the other three parameters.
The value of the neutron EDM in the MSSM 
is much less than the value of $10^{-27}$ indicated in \cite{INUI95},  because
values of the $A$ parameter as large as those used in ref.\cite{INUI95},  give
problems with colour or charge  breaking minima, or do not lead to a solution
for $\mu$ and $B$  on minimisation. We find that the expected range for the EDM
is  therefore     
\begin{equation}
10^{-33}<|d_n |< 10^{-29} {e\;\mbox{cm}} .
\end{equation}

It is expected that future developments will push the experimental  bound on
the neutron EDM down from eq.(\ref{edm-2}) to $O(10^{-28})  {e\;\mbox{cm}} $
but, as the present calculations indicate, will still not
provide  a test of the constrained MSSM\footnote{We thank K.~Green
for discussions}.  A more  remote possibility is the direct measurement of the
$t$-quark EDM  using 
either $t\bar{t}$ decay correlations in $e^{+}e^{-} \rightarrow t\bar{t}$
\cite{BNOS92} or $t\bar{t}$ production via photon-photon fusion using
linearly polarised photons generated by Compton 
back-scattering of laser light on electron or positron beams of  linear
$e^{+}e^{-}$ or $e^{-}e^{-}$ colliders \cite{CH95}. In this analysis, we found
that the $t$-quark EDM is usually larger than the neutron EDM by a factor of
3--5, a slight improvement but still unlikely to be measured in the  
forseeable future, at least in the MSSM.
                  
\newpage

{\bf Acknowledgments}

\noindent
IBW would like to thank the University of Bristol for the award of a   
Benjamin Meaker Visiting Professorship for the period during which 
this research was undertaken. We would like to thank P.~Gondolo, K.~Green, 
I.~Jack, P.~Ohmann, C.~Kolda and R.~G.~Roberts for discussions.

\newpage
{\bf \Large{Figure Captions}}
\begin{itemize}
\item[Figure 1~:] Quark self energy diagrams involving (a) gluino and 
(b) chargino exchange. 
\item[Figure 2~:] SUSY contribution to quark EDM from chargino-photon 
coupling. 
\item[Figure 3~:] The modulus of Neutron EDM (in units of $10^{-33}\;e$ cm) as a 
function of
$\tan \beta $ for random choices of $m_0,\;m_{1/2}$ and $|A|$ in the 
range (0, 1) TeV.
\end{itemize}
\vfill
\end{document}